\documentclass[aps,pra,reprint,groupedaddress]{revtex4-1}

\usepackage{soul,color}

\usepackage{amsmath,bm}
\usepackage{graphicx}
\usepackage[raggedright,hang,nooneline]{subfigure}
\usepackage{float}

\begin{document}


\title{Distribution of laser shot noise energy delivered to a levitated nanoparticle}


\author{T. Seberson$^{1}$ and F. Robicheaux$^{1,2}$}
\affiliation{$^{1}$Department of Physics and Astronomy, Purdue University, West Lafayette, Indiana 47907, USA}
\affiliation{$^{2}$Purdue Quantum Science and Engineering Institute, Purdue University, West Lafayette, Indiana 47907, USA}


\date{\today}

\begin{abstract}
This paper quantifies the rate at which laser shot noise energy is delivered to a nanoparticle for the various scenarios commonly encountered in levitated optomechanics. While previous articles have the same form and dependencies, the proportionality constants often differ in the literature. This paper resolves these discrepancies. The rate at which energy is delivered to an optically trapped particle's respective degrees of freedom depends on the radiation pattern of scattered light as well as the direction of laser propagation. For a traveling plane wave with linearly polarized light, in the Rayleigh regime this leads the translational shot noise heating rate to be proportional to $1/10$ of the total rate in the laser polarization direction, $7/10$ in the laser propagation direction, and $2/10$ in the direction perpendicular to both. Analytical expressions for the shot noise heating rate are provided in the Rayleigh limit as well as numerical calculations for particles in the Mie regime for silica and diamond. For completeness, numerical calculations of the shot noise heating for silica Mie particles at the focus of a strongly focused laser beam are calculated for varying numerical aperture and common laser wavelengths. Both numerical calculations show that the Rayleigh expression generally gives an overestimate of the shot noise heating especially for larger radii, but is still a good approximation even for incident focal fields. The exception to the relative decrease is when a Mie resonance is reached which was found for diamond. Lastly, Rayleigh expressions for the rotational shot noise heating for a symmetric top-like particle for linear, elliptically, and unpolarized light are also provided. %
\end{abstract}


\maketitle

\section{Introduction}\label{intro}
As the field of levitated optomechanics nears an era where nanoparticles are able to be cooled near their motional ground state \cite{PhysRevLett.122.223601,PhysRevLett.122.123602,Delic2020,PhysRevLett.122.123601}, the effects of heating and noise in these systems becomes essential to understand and quantify. Besides experimental limitations such as imperfect detection efficiency and phase noise, a standard levitated dielectric particle trapped in the focus of a laser beam experiences heating and damping from the surrounding gas as well as laser shot noise. The scattering of gas molecules has been studied extensively \cite{Epstein1924,doi:10.1002/andp.201200232,Millen_2020} and may become a negligible effect for sufficiently low pressures. However, for optically trapped particles, laser shot noise is an inescapable factor of consideration for particles with low motional occupation numbers. 

Previous studies and reviews have included translational shot noise heating in their analyses \cite{e20050326,Neukirch_2014,PhysRevA.100.013805,Jain2016, PhysRevLett.123.153601,Gieseler2012,Deli__2020,PhysRevA.95.053421,Chang1005}, however there were often inconsistencies. In this paper we give a detailed description of the phenomenon of laser shot noise heating for various scenarios in levitated optomechanics. The main focus is to quantify the amount of translational energy delivered to a harmonically trapped nanoparticle's degrees of freedom due to the scattering of photons. The translational calculations performed here are based on the derivation in Ref. \cite{Itano1982}, but the results may also be obtained using a full quantum treatment \cite{PhysRevA.21.1606,PhysRevA.86.013802}. The difference from Ref. \cite{Itano1982} is that the scattering object is considered to be a nanoparticle instead of an atom. 

The most common expression used for the shot noise heating rate is that which is derived for a plane wave incident upon a Rayleigh particle due to its ease of calculation and range of applicability. Whether this expression is a valid estimation for finite sized particles with incident plane waves or a traveling focused Gaussian beam is an outstanding question. In this paper the shot noise heating rate for each translational degree of freedom is calculated numerically for spherical Mie particles up to radii of $250\,{\rm nm}$. The calculations were performed for an incident plane wave and for a focused Gaussian laser beam for commonly used laser wavelengths. Mie scattering calculations were performed for silica and diamond nanospheres. A discrete-dipole approximation method was used to numerically evaluate the scattered fields for the incident focused Gaussian laser beam. The calculations show that the Rayleigh expression generally gives an overestimate of the shot noise heating especially for larger radii. The exception being for parameters that lead to a Mie resonance. For finite sized silica nanospheres, the values of the shot noise obtained for an incident focused Gaussian beam are within an order of magnitude of the Rayleigh expression for each degree of freedom. The discrepancy increases as the radius of the particle and/or the numerical aperture of the lens increases.  

For non-spherical particles, there is shot noise heating in the rotational degrees of freedom as well. The procedure outlined in Refs. \cite{Papendell_2017,Stickler2016} is correct, but the expressions for the diffusion constants appear to be calculated for a particle illuminated with unpolarized light with random propagation direction. Additionally, the original calculations performed in Ref. \cite{PhysRevA.95.053421} are reduced by a factor of $1/2$ from the actual value. For completeness, expressions for the rotational shot noise heating rate are provided for different laser polarizations. 

This paper is organized as follows. In Sec. \ref{trans_shot_noise} the amount of translational energy delivered to an optically trapped particle is provided. Its subsections investigate the rate at which the energy is delivered to the particle in each degree of freedom for both Rayleigh scattering and Mie scattering for incident plane waves and a focused Gaussian beam. In Sec. \ref{rot_heat}, the Rayleigh expressions for the rotational shot noise heating for a symmetric top-like particle for linear, unpolarized, and elliptically polarized light are given.

\section{Translational shot noise}\label{trans_shot_noise}
The derivation of the translational shot noise follows from Ref. \cite{Itano1982} which was originally calculated for atoms in a far red-detuned dipole trap. The calculations in this section are based on semi-classical ideas. The quantum calculations yield identical results and may be found in Appendix \ref{AppendixA}.

Consider the scattering of a single photon with wavenumber $\vec{k}_i=k\hat{k}_i$ off a particle with initial momentum $m\vec{v}_i$. In the regime $|\vec{v}_i| \ll c$, after the scattering event the photon has final momentum $\hbar \vec{k}_f \approx \left( \hbar k\right) \hat{k}_f$ and the particle has final momentum $m\vec{v}_f$. From conservation of momentum,
\begin{align}
\vec{v}_f=\vec{v}_i + \frac{\hbar}{m} \left( \vec{k}_i-\vec{k}_{f} \right). \label{eq1}
\end{align}
The change in energy for the component $j = \left( x, y ,z \right)$ is
\begin{align}
\begin{split}
\Delta E_{j} &= \frac{1}{2}m\left( v^{2}_{fj} - v^{2}_{ij} \right) \\  \label{eq2}
 &= \bm{\varepsilon}\left( \hat{k}^{2}_{ij} + \hat{k}^{2}_{fj} - 2\hat{k}_{ij}\hat{k}_{fj} \right) + \hbar k\left( \hat{k}_{ij} - \hat{k}_{fj} \right)v_{ij} , \\
\end{split}
\end{align}
with $\bm{\varepsilon} =  \frac{\hbar^{2}k^{2}}{2m}$. The change in energy of the particle then depends on the initial photon propagation direction $\hat{k}_{i}$ as well as the scattered direction $\hat{k}_{f}$. The above equations are accurate for a free particle, but are also valid for harmonically bound particles if the scattering takes place on time scales much shorter than the oscillation frequency. 

The probability for the photon to scatter into a solid angle $d\Omega$ is 
\begin{equation}
P(\hat{k}_f)d\Omega = \frac{1}{\sigma}\bigg(\frac{d\sigma}{d\Omega}\bigg)d\Omega , \label{eq3}
\end{equation}
where $d\sigma/d\Omega$ is the differential scattering cross section for the particle and $\int P(\hat{k}_f)d\Omega =1$. The average change in energy $\langle\Delta E_j\rangle$ of the particle following the scattering event is found through
\begin{equation}
\langle \Delta E_j\rangle = \int_{\Omega} P(\hat{k}_f)\Delta E_j d\Omega .   \label{eq4}
\end{equation}
To evaluate Eq. (\ref{eq4}), the particle is taken to be a sphere that is oscillating in a harmonic potential. The incident photon is traveling in the $\hat{k}_{i} = \hat{z}$ direction and polarized in the $\hat{E}_{\rm inc}=\hat{x}$ direction. Immediately, the contribution from the last term in Eq. (\ref{eq2}) goes to zero, $\langle\hbar k\left( \hat{k}_{ij} - \hat{k}_{fj} \right)v_{ij}\rangle =0$, since $\langle \vec{v}\rangle=0$ for harmonic oscillation.  Looking at the change in energy in each direction explicitly, Eq. (\ref{eq4}) is rewritten as
\begin{subequations}  \label{eq6}
\begin{align}
\begin{split} 
\langle \Delta E_x\rangle &= \bm{\varepsilon}\int_{\Omega} P(\hat{k}_f) \left(\sin\theta\cos\phi  \right)^2 d\Omega, \\ \label{eq6a}
\end{split} \\
\begin{split} 
\langle \Delta E_y\rangle &= \bm{\varepsilon}\int_{\Omega} P(\hat{k}_f) \left( \sin\theta\sin\phi  \right)^2  d\Omega, \\ \label{eq6b}
\end{split} \\
\begin{split} 
\langle \Delta E_z\rangle &= \bm{\varepsilon}\int_{\Omega} P(\hat{k}_f) \left(1  -  \cos\theta \right)^2  d\Omega, \\ \label{eq6c}
\end{split}
\end{align}
\end{subequations}
where spherical coordinates were used to define the outgoing wave, $\hat{k}_{fx} = \sin\theta\cos\phi$, $\hat{k}_{fy} = \sin\theta\sin\phi$, $\hat{k}_{fz} = \cos\theta$. To complete Eq. (\ref{eq6}) the differential scattering cross section for the particle must be determined. In the subsections below the energy delivered to a particle in the Rayleigh and Mie regimes are computed. Note that from Eq. (\ref{eq6}) the total energy delivered to a particle 
\begin{equation}
\langle\Delta E\rangle = \sum_j \langle\Delta E_j\rangle=2\bm{\varepsilon}\left(1 - \int_{\Omega}  P(\hat{k}_f)\cos\theta d\Omega \right), \label{eq7}
\end{equation}
is always greater than zero with a maximum of $4\bm{\varepsilon}$.

\subsection{Rayleigh scattering}\label{rayleigh}
For an incident monochromatic plane wave polarized in the $\hat{E}_{\rm inc}=\hat{x}$ direction, the differential scattering cross section for a dipole with moment $\vec{p}=\alpha\vec{E}_{\rm inc}$, index of refraction $n$, and radius $r$ in the Rayleigh regime $kr|n-1|\ll 1$ takes the form \cite{jackson_classical_1999} 
\begin{equation}
\bigg(\frac{d\sigma}{d\Omega}\bigg) = \bigg(\frac{k^{2}\alpha}{4\pi\epsilon_{0}} \bigg)^2 \sum_j |\hat{\xi}_j \cdot \hat{E}_{\rm inc}|^2 ,   \label{eq8}  
\end{equation}
yielding
\begin{equation}
P(\hat{k}_f) = \left( \frac{3}{8\pi} \right)\left[ \cos^{2}\theta \cos^{2}\phi +\sin^{2}\phi \right],   \label{eq9} 
\end{equation}
where $\epsilon_0$ is the permittivity of free space and $\hat{\xi}_j(\hat{k}_f)$ defines the two orthogonal polarization directions of the scattered light perpendicular to $\hat{k}_f$ so that  $\sum_j |\hat{\xi}_j(\hat{k}_f) \cdot \hat{E}_{\rm inc}|^2 = 1-|\hat{k}_f \cdot \hat{E}_{\rm inc}|^2$. Note that probability densities of the form Eq. (\ref{eq9}) have the property $P(\hat{k}_f) = P(-\hat{k}_f)$. 
\begin{figure*}[t]
    \includegraphics[width=0.99\textwidth]{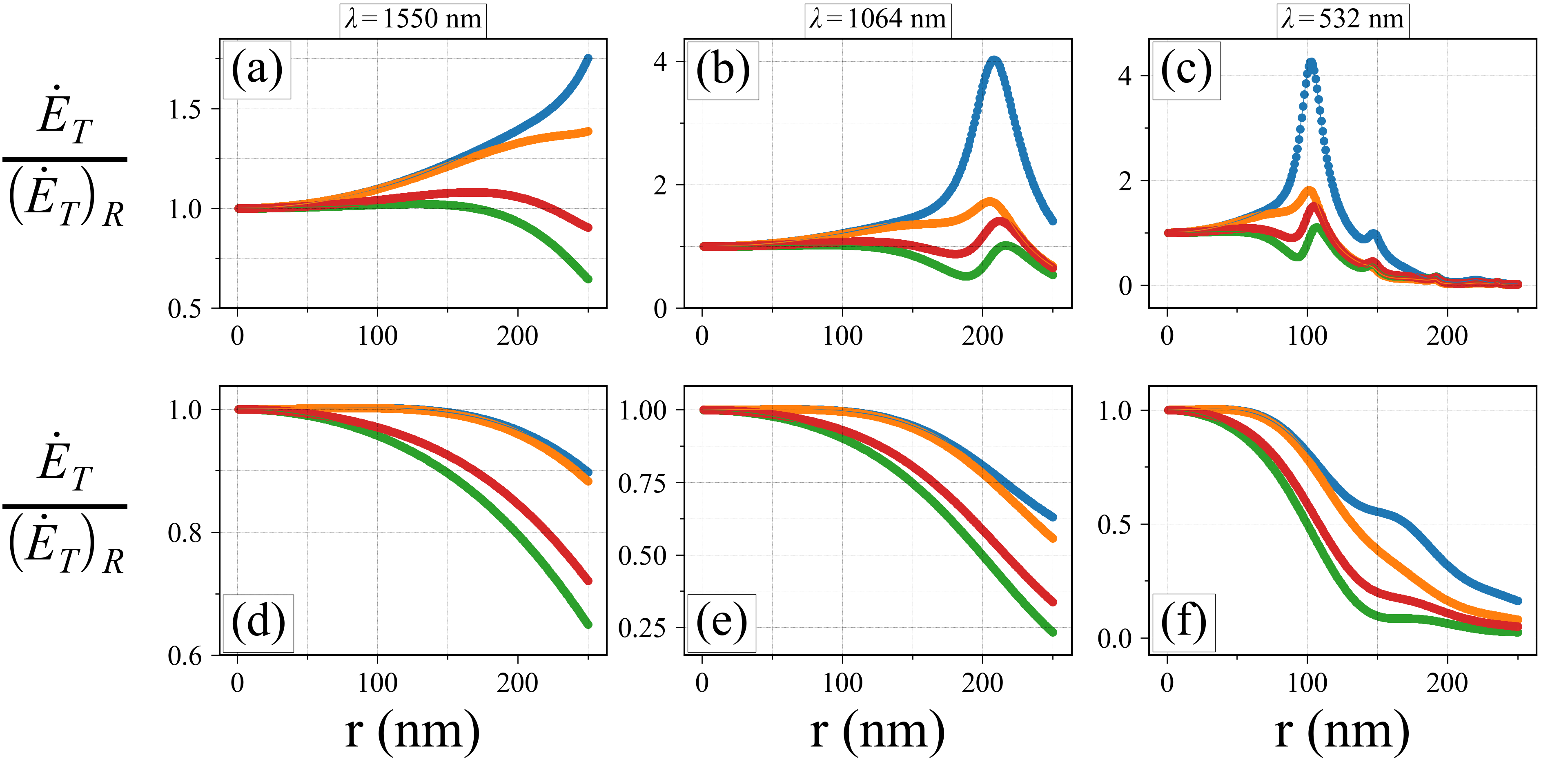} 
  \caption{ Shot noise heating rate for each degree of freedom relative to the rate in the Rayleigh limit for diamond at laser wavelength (a) $\lambda=1550 \,\rm nm$ (b) $\lambda=1064 \,\rm nm$ (c) $\lambda = 532 \,\rm nm$ and silica at wavelength (d) $\lambda=1550 \,\rm nm$ (e) $\lambda=1064 \,\rm nm$ (f) $\lambda = 532 \,\rm nm$. For each plot, the relative heating rate is shown for the $x$ (blue), $y$ (orange), and $z$ (green) degree of freedom, as well as the total heating rate (red). Here, the incident plane wave is polarized in the $x$ and traveling in the $z$ direction. The relative heating rate is reduced in each degree of freedom for silica. The non-linear behavior of Mie scattering can be seen near $r=200 \,\rm nm$ in diamond  with a resonance occurring at $\lambda=1064 \,\rm nm$. The index of refraction for silica is $n_s=1.45$ for $\lambda=1550 \,\rm nm$ and $\lambda=1064 \,\rm nm$ while $n_s=1.46$ for $\lambda=532 \,\rm nm$. The index of refraction for diamond is $n_d=2.39$ for $\lambda=1550 \,\rm nm$ and $\lambda=1064 \,\rm nm$ while $n_d=2.425$ for $\lambda=532 \,\rm nm$. } \label{fig:1064} 
\end{figure*}

Inserting Eq. (\ref{eq9}) into Eq. (\ref{eq6}) gives the distribution of energy delivered to the particle
\begin{equation} \label{eq10}
\bigg(  \langle  \Delta E_x \rangle_R , \langle  \Delta E_y \rangle_R , \langle \Delta E_z \rangle_R \bigg)  = \bm{\varepsilon}\left(  \frac{1}{5} , \frac{2}{5} , 1+\frac{2}{5} \right),  
\end{equation}
where the subscript $R$ refers to Rayleigh scattering. Thus, the scattering of one photon in the Rayleigh limit increases the particle's total energy by $\langle\Delta E\rangle_R = \sum_j \langle\Delta E_j\rangle_R=2\bm{\varepsilon}$ half the maximum amount possible and in different proportions in each direction. It is this result (Eq. \ref{eq10}), that has been missing in the expressions for the shot noise in much of the literature. The particle gains $7/10$ of the total energy in the direction of photon propagation, $1/10$ of the total energy in the photon polarization direction, and $2/10$ in the remaining direction. The contribution that eluded many of the previous works was the $\hat{k}^{2}_{ij}$ term in Eq. \eqref{eq2}. This term gives a change in energy of $1\bm{\varepsilon}$ in the direction of photon propagation, as seen in Eq. \eqref{eq10}. For unpolarized light, the total energy increase is the same as for linearly polarized light $\langle\Delta E\rangle_R=2\bm{\varepsilon}$. However, the energy is distributed as $14/20$ of the total energy in the direction of photon propagation and $3/20$ of the total energy in each of the directions perpendicular to the photon propagation direction. 

The previous calculation shows how the energy is distributed to each degree of freedom. The average rate at which energy is being delivered to these degrees of freedom $\dot{E}_T$ (shot noise heating rate) is the change in energy per scattering event multiplied by the scattering rate. The scattering rate is the number of incident photons per unit area per unit time, $J_p=I_0/\hbar\omega$, times the scattering cross section, $\sigma$,
\begin{equation}
\dot{E}_{T_R} = \langle\Delta E\rangle_R J_p \sigma  .  \label{eq11}   
\end{equation}
For Rayleigh particles, $\sigma_R = \left(\frac{8\pi}{3}\right)\left(\frac{\alpha k^2}{4\pi\epsilon_0}\right)^2$. $\dot{E}_{T_{R}}$ is the total translational energy gained per second due to shot noise. This is written in many forms in the literature, but often in terms of the scattered power $\dot{E}_{T_{R}} =\hbar\omega_0 P_{\rm scatt}/ mc^2 $ with $\omega_0=ck$ the frequency of the laser. Note that this calculation is different from that which calculates the average radiation pressure force in the axial direction. Nor is this a calculation of radiation damping due to scattered light, where we have seen the expression for the two different effects confused in the literature \cite{PhysRevA.96.032108,e20050326}. 

If the particle oscillates at frequency $\omega_j$ in the $j$th direction, each degree of freedom's occupation number increases at a rate of 
\begin{equation}
\Gamma_{x} = \frac{1}{10} \frac{\dot{E}_{T_R}}{\hbar\omega_x}, \quad \Gamma_y = \frac{2}{10} \frac{\dot{E}_{T_R}}{\hbar\omega_y} , \quad \Gamma_z = \frac{7}{10} \frac{\dot{E}_{T_R}}{\hbar\omega_z}.  \label{eq12}
\end{equation}
While the rates in Eq. (\ref{eq12}) have the same form as the expressions seen previously \cite{Neukirch_2014,PhysRevA.100.013805,Jain2016,PhysRevLett.123.153601,Gieseler2012,Deli__2020,PhysRevA.95.053421}, the factors in front of them are different. These factors may also become important when considering experiments which hope to measure heating and/or noise for clues to new physics \cite{PhysRevA.94.010104} or attempting to cool to the motional ground state \cite{PhysRevLett.122.123602,Delic2020}. 

Our derivation of the rate that shot noise increases the energy of a nanoparticle agrees with that in Ref. \cite{PhysRevA.100.013805}. Reference \cite{PhysRevA.100.013805} defines "the recoil heating rate", Eq. (B33), to be 1/2 the value we obtain for the shot noise heating rate. However, in their definition, the rate that the energy increases is $2\times$ the value in Eq. (B33) in their paper, leading to agreement between our derivations. Reference \cite{Jain2016} supplied an expression for and measured the shot noise heating rate. Their expression for the two directions perpendicular to the laser propagation direction, $\hat{k}=\hat{z}$, are correct, while the total energy and proportionality constant for the $z$ degree of freedom is not. The measured value for the degree of freedom perpendicular to both the laser propagation and polarization directions was reported to be within error bars, while the values in the remaining directions were not reported. 

\subsection{Mie scattering\label{mie}}
The analytical expressions in the previous subsection are valid for small particles $kr|n-1|\ll 1$. For particles outside the Rayleigh regime $kr|n-1|\sim 1$ the differential scattering cross section in Eq. \eqref{eq8} and therefore Eq. (\ref{eq9}) breaks down and Mie scattering \cite{Wriedt2012} must be used to calculate the shot noise heating. Figure \ref{fig:1064} plots the translational shot noise heating rate $\dot{E}_{T}$ for each degree of freedom for varying particle radii using analytical Mie formulae. Specifically, $\dot{E}_{Tj}/J_p=\langle\Delta E_j \rangle\sigma $ for each degree of freedom is found by numerically calculating the differential scattering cross section and numerically integrating Eqs. \eqref{eq6}. These quantities are then divided by their respective Rayleigh expression, $(\dot{E}_{Tj})_R/J_p=\langle\Delta E_j \rangle_R \sigma_R $. 

For particle sizes $r\leq 50 \,\rm nm$, Fig. \ref{fig:1064} shows that the total shot noise heating rate may still be approximated as the Rayleigh expression to better than $10\%$ error for both silica ($n_s=1.45$) and diamond ($n_d=2.39$) at a wavelength of $\lambda=1064 \,\rm nm$ or $\lambda=1550 \,\rm nm$. For larger particle sizes, the non-sextic behavior of the differential scattering cross section with respect to the radius becomes more apparent. As expected, deviations from the Rayleigh approximation become more significant as $kr|n-1|$ approaches unity with smaller wavelengths producing the most considerable change for both materials. For diamond, there is a resonance in the scattering near $r\sim 200 \,\rm nm$ for $\lambda=1064 \,\rm nm$ and a two order of magnitude suppression near $r\sim 250 \,\rm nm$ for $\lambda=532 \,\rm nm$.  
\begin{figure}[t]
    \includegraphics[width=0.45\textwidth]{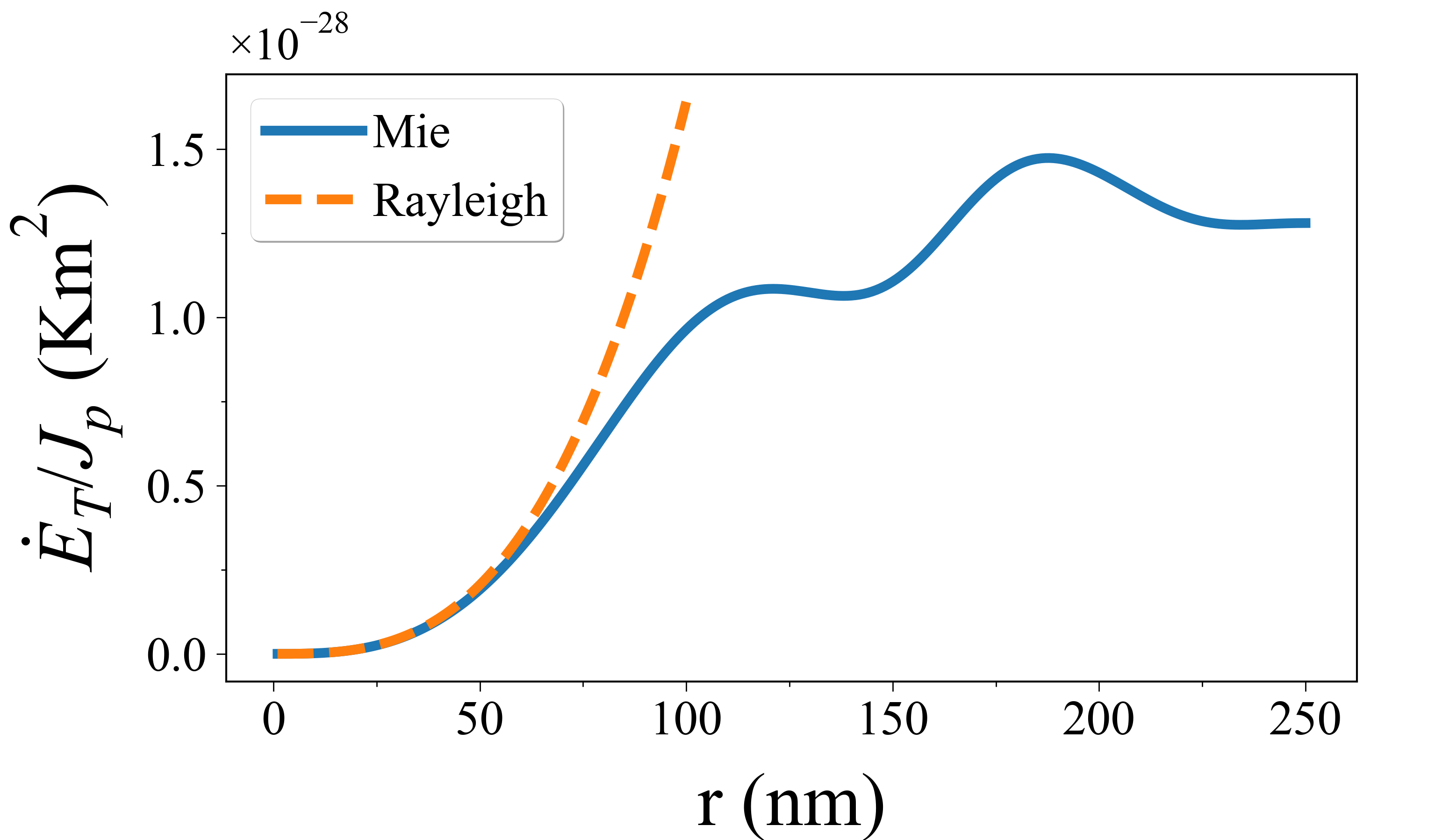} 
  \caption{ Comparison of the total shot noise heating rate using Mie and Rayleigh scattering for a silica nanoprticle ($n_s=1.46$ for $\lambda=532 \,\rm nm$) under plane wave illumination. For radii $r<50 \,\rm nm$ the Mie calculation shares the same $r^3$ dependence as the Rayleigh expression. For larger radii, the non-linear Mie calculation yields significantly less total heating.  } \label{fig:mie} 
\end{figure}

For $\lambda = 1550, 1064,$ and $532 \,\rm nm$, the shot noise heating rate for silica is decreased for each degree of freedom relative to the Rayleigh expression. From Fig. \ref{fig:1064}(f), the heating rate is almost an order of magnitude lower for $\lambda = 532 \,\rm nm$ near $r=200 \,\rm nm$. In Ref. \cite{2019arXiv190706046P} the laser shot noise was calculated for a $r=230 \,\rm nm$ silica nanoparticle illuminated with a $\lambda = 532 \,\rm nm$ laser. If their shot noise heating rate was calculated using the Rayleigh expression, Fig. \ref{fig:1064}(f) shows that that estimate should be reduced by $\sim 10$ times the calculated value, making laser shot noise an even less significant noise source for their experiment. Figure \ref{fig:mie} shows the explicit dependence of $r$ on the heating rate for silica at $\lambda = 532 \,\rm nm$ compared with the $r^3$ dependent Rayleigh expression. Here one can clearly see the range of accuracy of the Rayleigh approximation with the Mie calculation strongly deviating above $r\sim75 \,\rm nm$. 

\subsection{Focused Gaussian Beam\label{focused}}
\begin{figure*}[t]
    \includegraphics[width=0.99\textwidth]{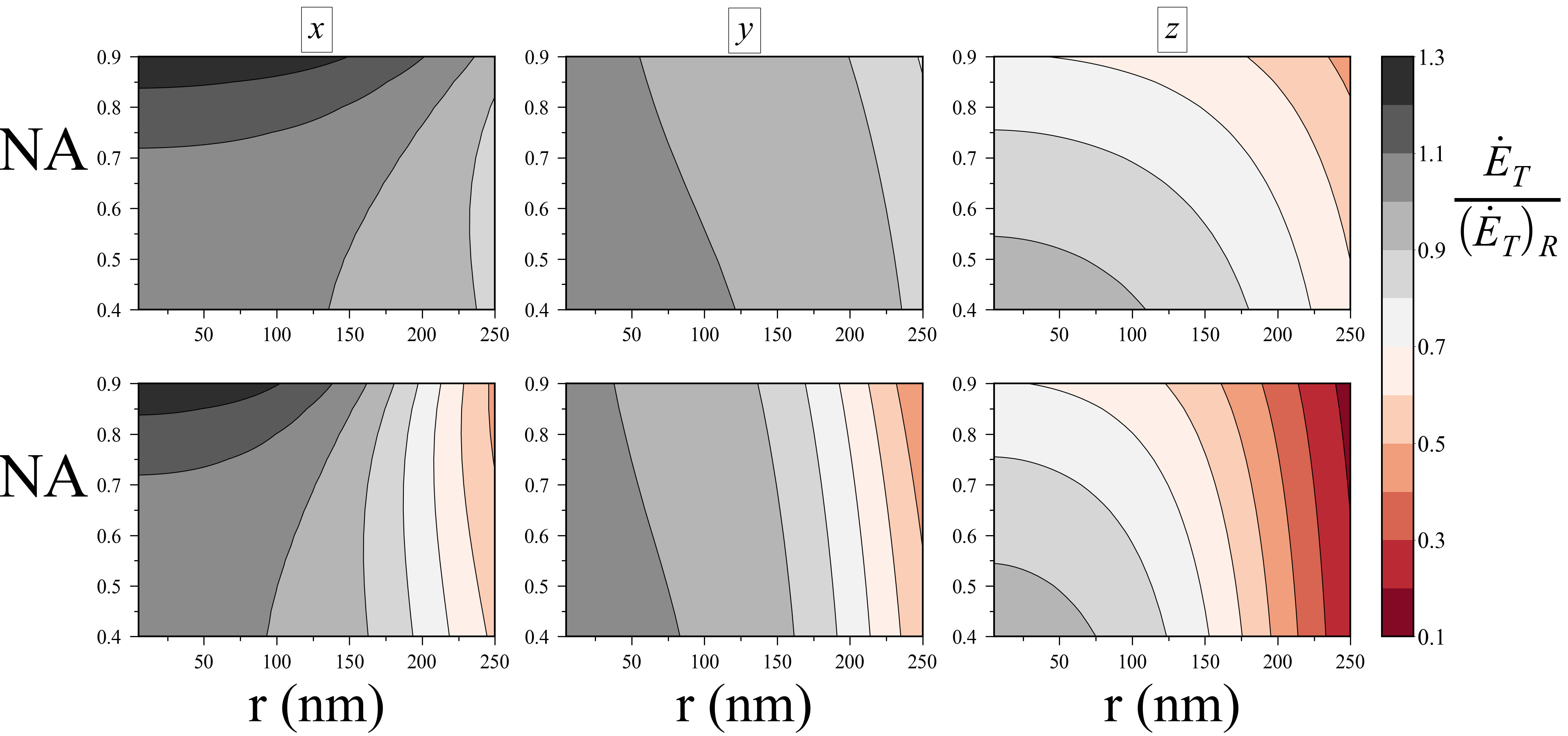} 
  \caption{ Contour surface plots of the shot noise heating rate for each degree of freedom relative to the rate in the Rayleigh limit for silica ($n=1.45$) at laser wavelengths $\lambda=1550 \,\rm nm$ (top row) and $\lambda=1064 \,\rm nm$ (bottom row). The heating rates for each degree of freedom $(x, y, z)$  are shown in the first, second, and third column respectively. The colorbar on the far right is a scale for the ratio between the numerically calculated shot noise heating, Eq. \eqref{eq115}, and the Rayleigh expression for that degree of freedom. As expected for small $\text{NA}$ and radius the shot noise in each degree of freedom agrees well with the Rayleigh expression.   } \label{fig:tweezer} 
\end{figure*}
Since an optically levitated nanoparticle is often trapped using a focused Gaussian laser beam it is practical to consider the shot noise heating due to such an incident wave. The electromagnetic fields of a Gaussian laser beam ((0,0) mode) focused through a lens with numerical aperture $\text{NA} = \sin\theta_{\rm max}$ can be expressed in cylindrical coordinates as \cite{novotny_hecht_2006,doi:10.1098/rspa.1959.0200}
\begin{subequations}  \label{eq113}
\begin{align}
\vec{E}(\rho,\phi,z) &= \frac{ikf}{2}E_0e^{-ikf} \begin{bmatrix}
I_{00} + I_{02}\cos 2\phi \\ I_{02}\sin 2\phi \\ -2iI_{01}\cos \phi \label{eq113a}
\end{bmatrix}, \\
\vec{H}(\rho,\phi,z) &= \frac{ikf}{2Z_0}E_0e^{-ikf} \begin{bmatrix}
I_{02}\sin 2\phi \\ I_{00} - I_{02}\cos 2\phi \\ -2iI_{01}\sin \phi    \label{eq113b}
\end{bmatrix} ,
\end{align}
\end{subequations}
where $Z_0$ is the impedance of free space, $f$ is the focal length of the lens, and the $I_{i,j}$ are integrals over the polar angle up to the extent of the lens, $\theta_{\rm max}$. The expressions for the integrals may be found in Appendix \ref{AppendixC} and we refer the reader to Ref. \cite{novotny_hecht_2006} for a detailed discussion of Eqs. \eqref{eq113}. 

Since the incident wave is no longer a plane wave, Eqs. \eqref{eq6} are not suitable for describing the energy transfer from the laser to the particle. Following a method similar to that of \cite{PhysRevA.86.013802} and \cite{PhysRevA.94.052109}, the shot noise heating rate for a general particle and incident wave can be written compactly as 
\begin{equation}
\dot{E}_{Tx_i} = J_p \bm{\varepsilon}  \int d^{2}\hat{k'}  \left\rvert \frac{\partial f(x_i,\vec{k},\vec{k'})}{\partial x_i}  \right\rvert^2_{x_i=0} , \label{eq115}  
\end{equation}
where $x_i = (x, y, z)$ and $f(x_i,\vec{k},\vec{k'})$ is the scattering amplitude. For an incident plane wave, $f(\vec{r},\vec{k},\vec{k'}) = e^{i\vec{k}\cdot \vec{r} } f(\vec{k},\vec{k'})e^{-i\vec{k'}\cdot \vec{r} } $, yielding Rayleigh shot noise, Eqs. \eqref{eqA5} in Appendix \ref{AppendixA}. 

To compute the shot noise heating rate using Eq. \eqref{eq115} for a sphere of radius $r$ with incident waves given by Eqs. \eqref{eq113} the scattered fields must be obtained. One approach is to combine Mie theory with the highly focused fields which has been undertaken in Ref. \cite{Neto_2000} to evaluate optical forces. In the present paper, the scattered fields are computed numerically by employing the discrete-dipole approximation method (DDA) \cite{Draine:94}. In the DDA, the spherical particle is composed of $N$ discrete spherical dipoles each with polarizibility $\alpha$ and the internal fields of the dielectric are solved for self-consistently to retrieve the scattered fields outside the particle. Once the scattered fields are obtained the scattering amplitude can be determined. In the implementation of the DDA used for this paper, each dipole that composed the spherical particle had a polarizibility  $\alpha = 4\pi\epsilon_{0}R^{3}\left( \dfrac{\epsilon-1}{\epsilon+2} \right)$. 

The shot noise heating rates relative to the Rayleigh expression for various numerical apertures and particle radii are shown in Fig. \ref{fig:tweezer}. The calculations were performed for a spherical particle composed of silica, $n=1.45$, with its center of mass located at the focus, $\vec{r}_0 = \langle 0, 0, 0 \rangle$. The heating rates for each degree of freedom $(x, y, z)$ (first, second, and third column) were computed for two different laser wavelengths $1550\, \rm nm$ (top row) and $1064\, \rm nm$ (bottom row). The waist of the Gaussian laser beam just before entering the focusing lens was chosen to be twice the aperture radius of the lens, corresponding to a filling factor $f_0=2$ for all values in the figure (see Appendix \ref{AppendixC}). 

As the radius of the particle increases, the shot noise heating decreases relative to the Rayleigh expression for all NA. As the radius increases, the intensity per volume decreases. Rayleigh expressions assume uniform incident plane waves while a tweezer has a Gaussian-like spot size. 

When the numerical aperture increases, the $\hat{z}$ component of the beam is more prominent, resulting in a polarization mainly in the $x-z$ plane (the $y$ component is negligible). It is plausible that since shot noise heating is smallest in the degree of freedom associated with the polarization direction, the shot noise in the axial degree of freedom decreases while heating increases in the $x$ degree of freedom. 

As the beam becomes more focused, the beam diverges more strongly as it exits the focal region. As opposed to a plane wave which always propagates in the $\hat{k}_i=\hat{z}$ direction, the incident wave through the particle due to a focused beam has propagation components in the $\hat{x}$ and $\hat{y}$ directions as well, decreasing the shot noise in the $\hat{z}$ degree of freedom. This comes from the main factor of discussion in Sec. \ref{rayleigh}, the $\hat{k}^{2}_{ij}$ term in Eq. \eqref{eq2} which gave a change in energy of $1\bm{\varepsilon}$ in the direction of photon propagation for Rayleigh scattering in Eq. \eqref{eq10}. This influence thus decreases for focused beams. In fact, Ref. \cite{PhysRevA.100.043821} introduced a geometric factor $A \leq 1$ helping to explain this effect. The factor allows for an approximate evaluation of the energy delivered to the particle in the $z$ degree of freedom 
\begin{equation}
 \langle \Delta E_z \rangle_R  \approx \bm{\varepsilon} \left(   A^2 + \frac{2}{5} \right) . \label{eq116}
\end{equation}
The geometrical factor is a result of a first order expansion of Eqs. \eqref{eq113} valid for particles small compared with the wavelength. The expression for $A$ is a ratio of integrals \cite{PhysRevA.100.043821} and approximates to $A \approx 1 - \left( k z_R \right)^{-1}$ for small $\text{NA}$, where $z_R$ is the Rayleigh range of a paraxial Gaussian beam \cite{Seberson2019}. 

Insertion of Eq. \eqref{eq116} into Eq. \eqref{eq11} for the shot noise in the $z$ degree of freedom should be accurate for $\lambda \gg r$. Using the data obtained in Fig. \ref{fig:tweezer} for particles with $r = 5 \,\rm nm$, the shot noise agrees with the approximate evaluation using Eq. \eqref{eq116} to within $2\%$ for all $\text{NA}$ and we use this as a benchmark for the accuracy of our calculations using DDA. 

The reduction in shot noise energy delivered to the $z$ degree of freedom does not result in an increase in energy in the other degrees of freedom while the particle is situated at the origin. Since the particle is a sphere and the beam is symmetrical about the origin, the $\hat{x}$ and $\hat{y}$ components of the incident wavevectors on the particle in the $-\hat{z}$ half space are reflections of the $\hat{x}$ and $\hat{y}$ outgoing wavevectors in the $+\hat{z}$ halfspace, canceling the influence of the incident propagation direction on the shot noise in the $x$ and $y$ degrees of freedom. 

The overall magnitudes for the shot noise in each degree of freedom in Fig. \ref{fig:tweezer} are within an order of magnitude of the Rayleigh expression up to a radius of $r=250 \,\rm nm$. Fortunately for experimentalists attempting to reach the motional quantum ground state, the values decrease as the radius increases for all degrees of freedom. This allows the Rayleigh expression for the shot noise heating rate to be used as an upper bound for calculations and a good approximation for all $\text{NA}$. 

A natural next question is how the shot noise would be distributed for a particle in a standing Gaussian wave, the situation of consideration for particles trapped in a driven cavity. Differing from tweezer traps, cavity traps typically have very large beam waists  $\sim 40\,\mu\text{m}$ \cite{PhysRevLett.122.123602,Kiesel14180} and therefore the radial geometry of the beam is well approximated as a symmetric Gaussian. Over the range of a nanoparticle $r \sim 100 \,\rm nm$ the field is essentially uniform and the beam can be approximated as an incident plane wave. In the axial direction the field dependence is of the form $\sim \cos kz$. The particle is placed in two common locations $z=0$ for cavity trapping and $z=\lambda/8$ for maximal cavity coupling. The latter situation can be achieved by using a separate tweezing laser to place the particle at that location \cite{PhysRevLett.122.123602,Delic2020,PhysRevLett.122.123601}. The following discussion is with reference to the shot noise from the cavity photons solely. The shot noise heating rate for a silica Mie particle of varying radius located at $z=\lambda/8$ has been calculated analytically in Ref. \cite{PhysRevA.86.013802} and numerically for a cavity with $1064 \,\rm nm$ wavelength and $26\,\mu\text{m}$ waist. Up to $\approx 250 \,\rm nm$ in radius, Fig. 4 in Ref. \cite{PhysRevA.86.013802} shows the shot noise increasing as the particle size increases. Our calculations confirm this result. However, it should not follow the traditional Rayleigh $r^3$ dependence exactly. Owing to the large beam waist, the situation is similar to that in Fig. \ref{fig:1064}(e) in the present paper.

We would like to add that at $z=0$, the dependence on the initial photon propagation direction $\hat{k}^{2}_{ij}$ vanishes giving $A=0$ in Eq. \eqref{eq116}. The amount of shot noise delivered to the axial degree of freedom, $z$, is then equal to the amount delivered to the degree of freedom orthogonal to both the polarization and axial directions, $y$. However, at $z=\lambda/8$ the $\hat{k}^{2}_{ij}$ contribution returns, giving Eq. \eqref{eq10} for Rayleigh particles.

\section{Rotational shot noise}\label{rot_heat}
Rotational diffusion constants were calculated in Refs. \cite{Papendell_2017,Stickler2016}, but for incident light unpolarized in all three directions. For particles in levitated optomechanics, the particles are typically illuminated with linear, elliptical, or circular polarization. The rotational shot noise heating rate given in Ref. \cite{PhysRevA.95.053421} is reduced by a factor of $1/2$ from the actual value given (Eq. (\ref{eq14})). For these reasons, expressions for the rotational shot noise heating rate are provided below for a symmetric top-like particle in the Rayleigh limit. The derivation of the rates below closely follows that of Ref. \cite{PhysRevA.95.053421} and may be found in Appendix \ref{AppendixB}.

For a symmetric top-like particle \cite{Seberson2019} illuminated by a linearly polarized laser with $\vec{E}_{\rm inc}=E_{0} \exp(ikz) \hat{x}$, the particle's symmetry axis will tend to align near the polarization axis. In the limit of small angle oscillations, the total rotational shot noise heating rate is
\begin{align}\begin{split}
\dot{E}_{R_{\parallel}} &= J_p \left( \frac{16\pi}{3} \right)  \left( \frac{k^2}{4\pi \epsilon_0} \right)^2 \left( \alpha_{z} - \alpha_{x} \right)^2 \left(\frac{\hbar^2}{2I_{x}}\right),  \label{eq14} \\
\end{split}
\end{align}
where $I_x$ is the moment of inertia perpendicular to the symmetry axis and $\alpha_z$, $\alpha_x$ are the polarizabilites parallel and perpendicular to the symmetry axis, respectively. The amount of shot noise delivered to each Euler angle $\alpha,\beta,\gamma$ in the $z$-$y'$-$z''$ convention \cite{Seberson2019} is 
\begin{equation}
\left(\dot{E}_{\alpha}, \dot{E}_{\beta} , \dot{E}_{\gamma}\right)_{\parallel} = \frac{\dot{E}_{R_{\parallel}}}{2} \left( 1 ,  1, 0 \right). \label{eq15}
\end{equation}
If the particle's symmetry axis is orthogonal to both the laser polarization and laser propagation direction, the energy is disributed as
\begin{equation}
\left(\dot{E}_{\alpha}, \dot{E}_{\beta} , \dot{E}_{\gamma}\right)_{\perp} = \frac{\dot{E}_{R_{\parallel}} }{2} \left(1,0, 0\right). \label{eq17}
\end{equation}
For unpolarized light, it is the average of Eqs. (\ref{eq15}) and (\ref{eq17}). As the angle $\alpha \rightarrow\pi/2$, the amount of shot noise delivered to the $\beta$ degree of freedom decreases. This can be understood from a decoherence/measurement perspective. As the nanoparticle's symmetry axis becomes orthogonal to the laser polarization and propagation directions, the laser can no longer provide information about the orientation of $\beta$; while $\alpha=\pi/2$, all angles $0\leq\beta\leq\pi$ look equivalent with respect to the laser polarization direction. This leaves $\beta$ to be immeasurable and is therefore immune to decoherence/heating. 

For elliptically polarized light defined by $\vec{E}_{inc}=E_{0}<\cos\psi,i\sin\psi,0>$, if the ellipticity is weak so that the particle's symmetry axis is primarily aligned along the $\hat{x}$ direction,
\begin{equation}
\dot{E}_{R_{e}} =\frac{\dot{E}_{R_{\parallel}} }{4}\left[ 4\cos^2\psi + 2\sin^2\psi \right], \label{eq18} 
\end{equation}
and
\begin{equation}
\left(\dot{E}_{\alpha}, \dot{E}_{\beta} , \dot{E}_{\gamma}\right) = \frac{\dot{E}_{R_{\parallel}} }{2}  \left(  1 , \cos^2\psi  , 0\right). \label{eq19}
\end{equation}
The expression for linearly polarized light is recovered for $\psi \rightarrow 0$. As the ellipticity of the beam increases, the amount of shot noise delivered to the $\beta$ degree of freedom decreases. This follows from a similar argument made for the case $\alpha=\pi/2$ above while considering the $y$ component of the polarization direction.

\section{\label{Conclusion}Conclusion}
This paper clarifies the rate at which energy is distributed to an optically trapped particle due to laser shot noise. Beginning from conservation of energy and momentum, the energy delivered to the particle is found to depend on the scattered radiation pattern as well as the incident wave's propagation direction. The energy in each degree of freedom increases per scattering event, but with different proportionality constants. These proportionality constants carry over to the average heating rate for each degree of freedom. Analytical expressions for the shot noise heating rate are provided in the Rayleigh limit as well as numerical calculations for particles in the Mie regime for silica and diamond. The shot noise heating in each degree of freedom is also computed for silica Mie particles at the focus of strong and weakly focused Gaussian beams. For finite sized silica nanospheres, the values of the shot noise obtained for an incident focused Gaussian beam are within an order of magnitude of the Rayleigh expression for each degree of freedom. The discrepancy increases as the radius of the particle and/or the numerical aperture of the lens increases. The Rayleigh expression may serve as a good approximation and upper bound as it is typically larger than the rates found for particles with finite radius.

\begin{acknowledgments}

This work was supported by the Office of Naval Research (ONR) Basic Research Challenge (BRC) under Grant No. N00014-18-1-2371.

\end{acknowledgments}

\appendix
\begin{widetext}
\section{\label{AppendixA}Quantum calculation of the translational shot noise heating rate from an incident plane wave}
Although the results are equivalent, the calculation of the shot noise may also be done quantum mechanically. The model used to calculate the shot noise describes particle state decoherence due to scattering events with photons \cite{SchlosshauerMaximilian2014Tqta}. The decoherence in the system state generates diffusion in momentum space which leads to heating. This method has already been used in Ref. \cite{PhysRevA.95.053421} to calculate the total translational shot noise heating rate, but the result in that reference is reduced from the actual result by a factor of $1/2$. Reference \cite{Stickler2016} provides the correct formula for computing the translational shot noise rate, but does not evaluate it.

The system will be the same as that considered in the main text with the incident plane wave propagating in the $\hat{z}$ and polarized in the $\hat{x}$ direction in the Rayleigh regime. The particle density matrix $\rho(\vec{r} , \vec{r'})$ is written in the position basis with $ \vec{r} = (x,y,z)$. The state $ \vec{r}$ refers to the system before a scattering event and the primed coordinates refer to the system following a scattering event. Neglecting the unitary part of the time evolution the translational master equation reads 
\begin{equation}
\partial_t\rho( \vec{r} ,  \vec{r'}) = -\Lambda( \vec{r} ,  \vec{r'}) \rho( \vec{r} ,  \vec{r'}), \label{eqA1}
\end{equation}
where the decoherence rate $\Lambda(\vec{r} , \vec{r}')$ to first order and neglecting cross terms which do not contribute is
\begin{equation}
\Lambda( \vec{r} ,  \vec{r'}) = D_x\left( x-x' \right)^2 + D_y\left( y-y' \right)^2  + D_z\left( z-z' \right)^2, \label{eqA2}
\end{equation}
but with separate diffusion constants 
\begin{align}
\begin{split}
D_j &= J_p \int d^{3}\vec{k}\mu (\vec{k})\int d^{2}\hat{k} | f(\vec{k},\vec{k}')|^2 \frac{k^2}{2}|\hat{k}-\hat{k}'_j|^2 \label{eqA3}
\end{split}
\end{align}
where $|f(\vec{k},\vec{k'})|$ is the scattering amplitude and $\mu (\vec{k}) = \delta(\vec{k}-\vec{k'})$ is the distribution of the laser which may safely be taken to be a delta function. The shot noise heating rate may be calculated through 
\begin{equation}
\dot{E}_{T} = \frac{d}{dt} \langle\mathbf{H}\rangle = \text{Tr}(\mathbf{K} \partial_t \rho ), \label{eqA4}
\end{equation}
with $\mathbf{H} = \mathbf{K} + \mathbf{U}$ the translational Hamiltonian, $\mathbf{U}$ the potential energy whose term vanishes after carrying out the trace, and $\mathbf{K}=\mathbf{P}^2/2m$ is the kinetic energy. Combining Eqs. (\ref{eqA1}) through (\ref{eqA3}) and inserting into Eq. (\ref{eqA4})
\begin{subequations} \label{eqA5}
\begin{align}
\begin{split} 
\dot{E}_{Tx} &=  J_p \frac{\hbar^2 k^2}{2m}   \int d^{2}\hat{k} |f(\vec{k},\vec{k'})|^2 \left( \hat{k}'^2_x \right), \label{eqA6}  \\
\end{split} \\
\begin{split}
\dot{E}_{Ty} &=  J_p \frac{\hbar^2 k^2}{2m}  \int d^{2}\hat{k}  |f(\vec{k},\vec{k'})|^2 \left( \hat{k}'^2_y  \right),  \label{eqA7} \\
\end{split} \\
\begin{split}
\dot{E}_{Tz} &=  J_p \frac{\hbar^2 k^2}{2m}  \int d^{2}\hat{k}  |f(\vec{k},\vec{k'})|^2 \left( \hat{k}^2 + \hat{k}'^2_z - 2\hat{k}\hat{k}'_z \right). \label{eqA8} \\ 
\end{split}
\end{align}
\end{subequations}
%
Noting that $|f(\vec{k},\vec{k'})|^2= d\sigma/d\Omega$ and $ \bm{\varepsilon}= \frac{\hbar^2k^2}{2m}$, Eqs. (\ref{eqA5}) are equal to Eqs. (\ref{eq6}) and (\ref{eq11}) in Sec. \ref{trans_shot_noise}. 
%

\section{\label{AppendixB} Quantum calculation of the rotational shot noise heating rate }
The rotational heating rate may be calculated in a similar fashion to that in the previous section. Here, the shot noise heating rate due to elliptically polarized light will be presented. The rate for linearly polarized light is recovered by taking respective limits. The particle density operator in the orientational basis is $\rho(\Omega , \Omega')$ with $\Omega = (\alpha,\beta,\gamma)$ the Euler angles in the $z$-$y'$-$z''$ convention \cite{Seberson2019}. Let $\Omega$ refer to the system before a scattering event and primed coordinates refer to the system following a scattering event. The rotational master equation reads 
\begin{equation}
\frac{\partial}{\partial t}\rho(\Omega , \Omega') = -\Lambda(\Omega , \Omega') \rho(\Omega , \Omega'), \label{eqB1}
\end{equation}
where 
\begin{equation}
\Lambda = \frac{J_p}{2} \int d^{3}\vec{k} \mu (\vec{k}) \int d^2\hat{k'}|f_{(\Omega)}(\vec{k},\vec{k'}) - f_{(\Omega ')}(\vec{k},\vec{k'})|^2, \label{eqB2}
\end{equation}
is the decoherence rate, $f_{(\Omega)}(\vec{k},\vec{k'})$ is the scattering amplitude, and $\mu (\vec{k}) = \delta(\vec{k}-\vec{k'})$ is the distribution of the laser which will again be taken to be a delta function. The shot noise heating rate may be calculated through 
\begin{equation}
\dot{E}_{R} = \frac{d}{dt} \langle\mathbf{H}_R\rangle = \text{tr}(\mathbf{K}_{R}\frac{\partial}{\partial t} \boldsymbol{\rho} ), \label{eqB3}
\end{equation}
with $\mathbf{H}_R = \mathbf{K}_R + \mathbf{U}_R$ the rotational Hamiltonian, $\mathbf{U}_R$ the potential energy which has zero contribution in the above equation, and $\mathbf{K}_R$ is the rotational kinetic energy. In the $z$-$y'$-$z''$ convention, the rotational kinetic energy is \cite{Edmonds}
\begin{equation} \label{eqB4}
\mathbf{K}_R = -\frac{\hbar^2}{2I_{x}}\bigg[ \frac{\partial^2}{\partial \beta^2} + \cot(\beta)\frac{\partial}{\partial \beta} + \frac{1}{\sin^2(\beta)}\frac{\partial^2}{\partial \alpha^2} - \frac{2\cos(\beta)}{\sin^2(\beta)}\frac{\partial^2}{\partial \alpha \partial \gamma} +\left(\frac{I_x}{I_z} +\cot^2(\beta) \right) \frac{\partial^2}{\partial \gamma^2} \bigg].
\end{equation}
To evaluate Eq. (\ref{eqB3}), begin with the far field scattering amplitude for a point dipole
\begin{align}
f_{(\Omega)}(\vec{k},\vec{k'}) &= \left( \frac{k^2}{4\pi \epsilon_0 E_{0}} \right) \hat{\zeta}\cdot \vec{p}, \label{eqB5} 
\end{align}
with $\hat{\zeta}$ the polarization of the scattered light, $E_0$ the magnitude of the incident electric field, and $\vec{p} = \overset{\text{\tiny$\leftrightarrow$}}{R}^\dagger \overset{\text{\tiny$\leftrightarrow$}}{\alpha}_{0}  \overset{\text{\tiny$\leftrightarrow$}}{R} \vec{E}_{\rm inc}$ the polarization vector. For incident elliptical light defined by $\vec{E}_{\rm inc}=E_{0}<\cos\psi,i\sin\psi,0>$, 
\begin{align}
\begin{split}
\vec{p} &= \overset{\text{\tiny$\leftrightarrow$}}{R}^\dagger \overset{\text{\tiny$\leftrightarrow$}}{\alpha}_{0}  \overset{\text{\tiny$\leftrightarrow$}}{R} \vec{E}_{\rm inc} \\ \label{eqB6}
 &= E_{0}\begin{pmatrix}
 \cos\psi \Big[ \alpha_{x} + \big(\alpha_{z} - \alpha_{x}\big)\sin^{2}\beta\cos^{2}\alpha \Big] + i\sin\psi\Big[ \big(\alpha_{z} - \alpha_{x}\big)\sin^{2}\beta\cos\alpha\sin\alpha \Big] \\ 
 \cos\psi\Big[ \big(\alpha_{z} - \alpha_{x}\big)\sin^{2}\beta\cos\alpha\sin\alpha \Big]  + i\sin\psi\Big[ \alpha_{x} + \big(\alpha_{z} - \alpha_{x}\big)\sin^{2}\beta\sin^{2}\alpha \Big]\\ 
 \cos\psi\Big[ \big(\alpha_{z} - \alpha_{x}\big)\sin\beta\cos\beta\cos\alpha \Big]  + i\sin\psi\Big[ \big(\alpha_{z} - \alpha_{x}\big)\sin\beta\cos\beta\sin\alpha \Big]
 \end{pmatrix} \\
 & \equiv \big(\alpha_{z} - \alpha_{x}\big)E_{0} \begin{pmatrix}
 A_x + iB_x \\ A_y  + iB_y \\  A_z  + iB_z 
\end{pmatrix} 
+  
 \alpha_{x}E_{0}\begin{pmatrix}
\cos\psi  \\ i\sin\psi  \\ 0
\end{pmatrix}   ,
\end{split}
\end{align}
where the $A_j = A_j(\alpha ,\beta,\gamma)$ $\left(B_j = B_j(\alpha ,\beta ,\gamma )\right)$ are the real (imaginary) parts of the polarization vector component $j =(x,y,z)$.

For scattered light in the $\hat{r} = \langle \sin\theta\cos\phi, \sin\theta\sin\phi, \cos\theta \rangle$ direction in spherical coordinates, the outgoing polarization vector $\hat{\zeta}$ can take two directions $\hat{\theta} = \langle \cos\theta\cos\phi,$ $\cos\theta\sin\phi, -\cos\theta \rangle$ or $\hat{\phi} = \langle -\sin\phi, \cos\phi,0 \rangle$. As there is no preference for which polarization is chosen, the sum of the scattering amplitudes must be used in Eq. (\ref{eqB3}),
\begin{equation}
|f_{(\Omega)}(\vec{k},\vec{k'}) - f_{(\Omega ')}(\vec{k},\vec{k'})|^2 \rightarrow \left( \frac{k^2}{4\pi \epsilon_0 E_{0}} \right)^2\bigg(|\hat{\theta}\cdot \vec{p}  -   \hat{\theta}\cdot \vec{p'}|^2 +  |\hat{\phi}\cdot \vec{p}  -   \hat{\phi}\cdot \vec{p'}|^2 \bigg) . \label{eqB7}
\end{equation}
Performing the integrals in Eq. (\ref{eqB2}) gives the decoherence rate as
\begin{equation}
\Lambda = J_p\left(\frac{4\pi}{3}\right) \left( \frac{k^2}{4\pi \epsilon_0 } \right)^2\left(\alpha_{z} - \alpha_{x}\right)^2 \sum_j \left[ (A_j-A'_j)^2 + (B_j-B'_j)^2 \right], \label{eqB8}
\end{equation}
with $A'_j = A_j(\alpha ',\beta ',\gamma ')$ and $B'_j = B_j(\alpha ',\beta ',\gamma ')$. 

To calculate the shot noise using Eq. (\ref{eqB3}), note that the $\frac{\partial}{\partial \gamma}$ terms from Eq. (\ref{eqB4}) evaluate to zero as there is no $\gamma$ dependence in the polarization vector in Eq. (\ref{eqB6}). From here, the orientation of the nanoparticle relative to the incident polarization must be considered. For well librationally bound nanoparticles under weak elliptical polarization, the particle is undergoing oscillations for which the small angle approximation may be appropriately made, $\alpha \rightarrow 0 + \xi$ , $\beta \rightarrow \frac{\pi }{2} - \eta$. In this view, Eq. (\ref{eqB4}) may be rewritten
\begin{align}
\begin{split}
\mathbf{K}_R \rightarrow &  -\frac{\hbar^2}{2I_{x}}\bigg[ \frac{\partial^2}{\partial \beta^2} + \cot(\beta)\frac{\partial}{\partial \beta}  + \frac{1}{\sin^2\beta}\frac{\partial^2}{\partial \alpha^2}  \bigg] \\
 &\approx -\frac{\hbar^2}{2I_{x}}\bigg[  \frac{\partial^2}{\partial \eta^2} +  \frac{\partial^2}{\partial \xi^2}  \bigg] .\label{eqB9}
\end{split}
\end{align}
Inserting Eqs. (\ref{eqB8}) and (\ref{eqB9}) into Eq. (\ref{eqB4}) and taking the trace gives the total rotational shot noise for elliptically polarized light to second order
\begin{align}\begin{split}
\dot{E}_{R_{e}} &= J_p \left( \frac{4\pi}{3} \right)  \left( \frac{k^2}{4\pi \epsilon_0} \right)^2 \left( \alpha_{z} - \alpha_{x} \right)^2 \left(\frac{\hbar^2}{2I_{x}}\right) \left[ 4\cos^2\psi + 2\sin^2\psi \right] \label{eqB10} \\
 &=\frac{\dot{E}_{R_{\parallel}} }{4}\left[ 4\cos^2\psi + 2\sin^2\psi \right], \\
\end{split}
\end{align}
with the energy distributed as
\begin{equation}
\left(\dot{E}_{\alpha}, \dot{E}_{\beta} , \dot{E}_{\gamma}\right) = \frac{\dot{E}_{R_{\parallel}} }{2}  \left(  \cos^2\psi + \sin^2\psi , \cos^2\psi   , 0\right), \label{eqB11}
\end{equation}
which recovers Eqs. (\ref{eq18}) and (\ref{eq19}) in the main text. For linearly polarized light, $\psi = 0$ . To account for unpolarized light (Eqs. (\ref{eq17}) and (\ref{eq15})), the average of the results due to the two orientations $\alpha \rightarrow 0 + \xi$ , $\beta \rightarrow \frac{\pi }{2} - \eta$ and $\alpha \rightarrow \frac{\pi}{2} - \xi$ , $\beta \rightarrow \frac{\pi }{2} - \eta$ for linearly polarized light was computed. 

\section{\label{AppendixC} Focal fields }
The integral expressions for the quantities in Eqs. \eqref{eq113} are \cite{novotny_hecht_2006}
\begin{align}
I_{00} &= \int^{\theta_{\rm max}}_{0} f_{w}(\theta) \left( \cos\theta \right)^{1/2} \sin\theta \left(1+\cos\theta  \right) J_{0}(k\rho\sin\theta)e^{ikz\cos\theta}, \label{eqC1} \\
I_{01} &= \int^{\theta_{\rm max}}_{0} f_{w}(\theta) \left( \cos\theta \right)^{1/2} \sin^2\theta J_{1}(k\rho\sin\theta)e^{ikz\cos\theta}, \label{eqC2} \\
I_{02} &= \int^{\theta_{\rm max}}_{0} f_{w}(\theta) \left( \cos\theta \right)^{1/2} \sin\theta \left(1-\cos\theta  \right) J_{2}(k\rho\sin\theta)e^{ikz\cos\theta}, \label{eqC3}
\end{align}
where $J_{n}(x)$ is the Bessel function of order $n$, $f_{w}(\theta) = \exp \left( \sin^2\theta/ f^2_0\sin^2\theta_{max} \right)$ is the apodization function with the filling factor $f_0 = w_0/R_a$ is the ratio of the laser beam waist before the lens and the radius of the aperture ($f_0 =2$ for the figure in the main text). As in the main text, the numerical aperture $\text{NA} = \sin\theta_{\rm max}$.

\end{widetext}

\bibliographystyle{apsrev4-1}
\bibliography{shot_noise_paper}

\end{document}